\documentclass[aps,prc,twocolumn,groupedaddress,showpacs]{revtex4-1}

\usepackage{graphicx}

\begin{document}


\title{Intertwined effects of pairing and deformation
  on neutron halos in magnesium isotopes}


\author{H. Nakada}
\email[E-mail:\,\,]{nakada@faculty.chiba-u.jp}
\author{K. Takayama}

\affiliation{Department of Physics, Graduate School of Science,
 Chiba University,\\
Yayoi-cho 1-33, Inage, Chiba 263-8522, Japan}


\date{\today}

\begin{abstract}
Matter radii of the $^{34-40}$Mg nuclei are investigated
by self-consistent Hartree-Fock-Bogolyubov calculations
assuming the axial symmetry.
With the semi-realistic M3Y-P6 interaction,
the $N$-dependence of the matter radii observed in the experiments
is reproduced excellently.
Both the pairing and the deformation play significant roles
in an intertwined manner.
The $^{35}$Mg nucleus has a smaller radius than the neighboring even-$N$ nuclei,
which is attributed to its smaller deformation.
In contrast, a neutron halo is obtained in $^{37}$Mg.
We point out that, in contrast to the pairing anti-halo effect
that may operate on the even-$N$ nuclei,
the pair correlation enhances halos in odd-$N$ nuclei,
owing to the new mechanism which we call \textit{unpaired-particle haloing}.
The halo in $^{37}$Mg is predicted to have peanut shape in its intrinsic state,
reflecting $p$-wave contribution, as in $^{40}$Mg.
The $N$-dependence of the deformation is significant again,
by which the single-particle level dominated by the $p$-wave component
comes down.
\end{abstract}


\maketitle



\noindent\textit{Introduction.}

Nuclear radii are physical quantities which are accessible by experiments
and carry basic information of nuclear structure.
One of the exotic nuclear properties disclosed by radioactive beams
is halos near the neutron drip-line~\cite{ref:Rii94}.
Nuclear halos have often been detected via enhancement
of nuclear radii~\cite{ref:Tan85}.
While neutron halos had been observed in several light nuclei,
they come more difficult to access experimentally for heavier nuclei.
Relatively recently, significant enhancement in reaction cross sections
($\sigma_R$'s) was discovered in $^{37}$Mg~\cite{ref:Mg37exp,ref:Mg37exp2}.
This is a good evidence for a neutron halo,
because $\sigma_R$ (and the interaction cross section) well correlates
to the matter radius~\cite{ref:Tan85,ref:Kar75,ref:Wat14}.
$^{37}$Mg is the heaviest halo nucleus observed so far.
Moreover, as it is likely well-deformed,
$^{37}$Mg could exemplify a deformed halo~\cite{ref:MNA97,ref:Nak08}.

It is remarked as well that irregular neutron-number ($N$) dependence
of $\sigma_R$'s has been observed in this region,
which should be important to investigate the halos.
The measured reaction cross section in $^{35}$Mg, $\sigma_R(\mbox{$^{35}$Mg})$,
seems suppressed compared to those in $^{34,36}$Mg.
In contrast, $\sigma_R(\mbox{$^{37}$Mg})$ is larger
than $\sigma_R(\mbox{$^{36}$Mg})$ and $\sigma_R(\mbox{$^{38}$Mg})$.
It was insisted~\cite{ref:UHS17},
by phenomenological studies using the deformed Woods-Saxon (WS) potential,
that the staggering in $^{36-38}$Mg may be ascribed
to the pairing anti-halo effect~\cite{ref:BDP00} in $^{38}$Mg,
and to quenching of the pair correlation in $^{37}$Mg.
However, for deformed nuclei near the neutron drip-line,
effects of the deformation and the pairing
could be intertwined~\cite{ref:ZMRZ10}.
Moreover, suppression of $\sigma_R(\mbox{$^{35}$Mg})$
has not been explored sufficiently,
although some Skyrme Hartree-Fock (HF) plus BCS calculations
predicted relatively small matter radius~\cite{ref:GSAM14}.
It should also be kept in mind that
the pair correlation does not always reduce nuclear radii.
In Refs.~\cite{ref:CRM14,ref:ZCMR17}, it was shown that
the pair correlation can enhance halos via coupling to the continuum.
Capable of handling these effects in a single framework,
studies by fully self-consistent mean-field calculations
including both the pairing and the deformation are highly desired.
It is noted that fully self-consistent calculations
including the pair correlation are able to describe nuclear halos,
if an appropriate effective interaction or energy density functional
is applied~\cite{ref:MR96}.

In this work we have implemented
the axial Hartree-Fock-Bogolyubov (HFB) calculations in $^{34-38}$Mg,
applying the semi-realistic interaction M3Y-P6~\cite{ref:Nak13}.
This is the first application
of the M3Y-type semi-realistic interaction~\cite{ref:Nak03}
to deformed HFB calculations,
following the application to the deformed HF calculations
in Ref.~\cite{ref:SNM16}.
As shall be shown,
the $N$-dependence of the matter radii deduced from $\sigma_R$'s
is in excellent agreement with the data.
The results reveal that both the pairing and the deformation
play significant roles,
cooperatively in certain cases.
A new mechanism of nuclear haloing will be pointed out,
which works for odd-$N$ nuclei.

\noindent\textit{Calculations and theoretical aspects.}

For computation, we use the Gaussian expansion method,
as detailed and tested for the Gogny interaction in Ref.~\cite{ref:Nak08}.
The single-particle (s.p.) or the quasiparticle (q.p.) functions
are expressed by superposition of the spherical basis-functions,
whose radial parts are Gaussians with various ranges.
The results are insensitive to the range parameters
of the radial part, if they are appropriately chosen.
The basis functions are truncated by the orbital angular momentum $\ell$.
We here adopt $\ell_\mathrm{max}=7$,
where $\ell_\mathrm{max}$ is the maximum of $\ell$ in the model space.
It has been confirmed, up to normally deformed cases,
that error due to this truncation is not significant
if $\ell_\mathrm{max}$ is taken to be greater by four
than the highest $\ell$ of the occupied level at the spherical limit.
In investigating halos,
it is important to treat energy-dependent asymptotics
of s.p. wave functions at distance.
The present numerical method enables it
in the self-consistent mean-field theory in an efficient manner,
even with finite-range interactions~\cite{ref:NS02,ref:Nak06}.
Coupling to the continuum is taken into account
substantially~\cite{ref:Nak06,ref:NMYM09}.
The M3Y-P6 semi-realistic nucleonic interaction~\cite{ref:Nak13}
is adopted here,
which reasonably describes the pair correlations,
as well as the shell structure
and its dependence on $Z$ and $N$~\cite{ref:NSM13,ref:NS14},
in a wide range of the nuclear chart.
In odd-$N$ nuclei, the ground state (g.s.) should have
one q.p. on top of the HFB vacuum.
The blocking due to the q.p. is handled
by the interchange $(U,V)\,\leftrightarrow\,(V^\ast,U^\ast)$
for the q.p. state~\cite{ref:RS80},
and the energy minimization is carried out
under the equal-filling approximation~\cite{ref:EFA}.

Many of the Mg nuclei have been known to be well-deformed.
It has been predicted that the neutron-rich Mg isotopes
are also deformed~\cite{ref:Ham07}.
As well as halos, nuclear deformation influences nuclear radii.
However, it is not obvious whether and how we can separate these two effects.
We here assume the following relations~\cite{ref:Wat14},
\begin{eqnarray}
    \langle r^2\rangle &=& \frac{\bar{r}_0^2}{3}
    \bigg[\exp\Big(2\sqrt{\frac{5}{4\pi}}\beta\Big)
      +2\exp\Big(-\sqrt{\frac{5}{4\pi}}\beta\Big)\bigg]\,,\nonumber\\
    \frac{q_0}{A} &=& \frac{\bar{r}_0^2}{3}
    \bigg[2\exp\Big(2\sqrt{\frac{5}{4\pi}}\beta\Big)
      -2\exp\Big(-\sqrt{\frac{5}{4\pi}}\beta\Big)\bigg]\,,
  \label{eq:rbar-beta}
\end{eqnarray}
where $\langle r^2\rangle$ and $q_0$ are the mean-square matter radius
and the intrinsic mass quadrupole moment, respectively.
The parameters $\bar{r}_0$ and $\beta$ on the rhs correspond to
the root-mean-square (rms) matter radius at the spherical limit
and the deformation parameter.
Note that, for small $\beta$, we have~\cite{ref:BM1}
\begin{equation}
  \langle r^2\rangle \approx \bar{r}_0^2\Big(1+\frac{5}{4\pi}\beta^2\Big)\,,
\end{equation}
indicating enhancement of matter radius when the nucleus is deformed.
With Eq.~(\ref{eq:rbar-beta}), $\bar{r}_0$ and $\beta$ are extracted
from the HFB results of $\langle r^2\rangle$ and $q_0$
for individual nuclei.
Then effects of the deformation may be recognized from the $\beta$ values,
and other effects should be contained in $\bar{r}_0$.

Nuclear halos are identified from the density distribution $\rho(\mathbf{r})$
at large $r$,
for which the s.p. or q.p. wave functions near the Fermi level are responsible,
within the mean-field framework.
At sufficiently large $r$,
the nuclear force becomes negligible
and the HFB equation for a neutron q.p. is approximated by~\cite{ref:SkP}
\begin{eqnarray}
    \left(-\frac{1}{2M}\frac{d^2}{dr^2} - \lambda\right)\,[r\,U_k(\mathbf{r})]
    &\approx& \varepsilon_k [r\,U_k(\mathbf{r})]\,,\nonumber\\
    \left(-\frac{1}{2M}\frac{d^2}{dr^2} - \lambda\right)\,[r\,V_k(\mathbf{r})]
    &\approx& -\varepsilon_k [r\,V_k(\mathbf{r})]\,.
  \label{eq:HFBeq-asymp}
\end{eqnarray}
In Eq.~(\ref{eq:HFBeq-asymp}), $k$ represents an individual q.p. state,
whose energy is denoted by $\varepsilon_k\,(>0)$,
and $\lambda\,(<0)$ is the chemical potential.
$V_k$ ($U_k$) is  the wave function for the occupied (unoccupied) component
of $k$ at the HFB vacuum.
From Eq.~(\ref{eq:HFBeq-asymp}),
the following asymptotic forms are derived,
\begin{eqnarray}
 r\,V_k(\mathbf{r})&\approx& \exp(-\eta_{k+} r)\,,\nonumber \\
 r\,U_k(\mathbf{r})&\approx&\left\{\begin{array}{ccc}
 \exp(-\eta_{k-} r) & \mbox{for}& -|\lambda|+\varepsilon_k < 0 \\
 \cos(p_k r + \theta_k) & \mbox{for}& -|\lambda|+\varepsilon_k > 0 \end{array}
 \right.,
\label{eq:qpwf-asymp}\end{eqnarray}
where
$\eta_{k\pm} = \sqrt{2M(|\lambda|\pm\varepsilon_k)}$,
$p_k = \sqrt{2M(-|\lambda|+\varepsilon_k)}$
and $\theta_k$ is an appropriate real number.
Whereas we have neglected the centrifugal potential
in Eq.~(\ref{eq:HFBeq-asymp})
that is relevant to the spin-angular parts
of $U_k(\mathbf{r})$ and $V_k(\mathbf{r})$,
halos are contributed primarily
by the $s$- and $p$-wave components~\cite{ref:RJM92}.

The asymptotic behavior of $\rho(\mathbf{r})$
is derived from Eq.~(\ref{eq:qpwf-asymp})~\cite{ref:Nak06,ref:SkP}.
The density distribution is given by
\begin{equation} \rho(\mathbf{r}) = \sum_k \big|V_k (\mathbf{r})\big|^2\,,
\label{eq:Rmat3}\end{equation}
for an even-even nucleus.
Let us denote the smallest q.p. energy by $\varepsilon^\mathrm{min}$,
and define $\eta^\mathrm{min}_\pm=\sqrt{2M(|\lambda|\pm\varepsilon^\mathrm{min})}$,
correspondingly.
The asymptotic form of $\rho(\mathbf{r})$ is then obtained as
\begin{equation}
 r^2\rho(\mathbf{r})\approx \exp(-2\eta^\mathrm{min}_+ r)\,.
\label{eq:dns-asymp}\end{equation}
Since $\eta^\mathrm{min}_+>\sqrt{2M\varepsilon^\mathrm{min}}$
and $\varepsilon^\mathrm{min}$ is comparable to or larger than the pairing gap,
halos could be hindered,
apart from effects of coupling to the continuum~\cite{ref:CRM14,ref:ZCMR17}.
This is known as the pairing anti-halo effect~\cite{ref:BDP00}.
In contrast, the density distribution of an odd-$N$ nucleus is
\begin{equation} \rho(\mathbf{r})
  = \sum_{k\,(\ne k_1)} \big|V_k (\mathbf{r})\big|^2
  + \big|U_{k_1} (\mathbf{r})\big|^2\,,
\label{eq:Rmat4}\end{equation}
where $k_1$ stands for the q.p. state which is occupied in the g.s.,
usually satisfying $\varepsilon^\mathrm{min}=\varepsilon_{k_1}$.
This leads to the asymptotic behavior of $\rho(\mathbf{r})$ as
\begin{equation}
 r^2\rho(\mathbf{r})\approx \exp(-2\eta^\mathrm{min}_- r)\,,
\label{eq:dns-asymp2}\end{equation}
instead of Eq.~(\ref{eq:dns-asymp}).
Unlike the even-$N$ case,
$\rho(\mathbf{r})$ may decay very slowly for increasing $r$,
possibly producing a halo
if $\varepsilon^\mathrm{min}\approx|\lambda|$.
Remark that this mechanism,
which we shall call \textit{unpaired-particle haloing},
works even for non-vanishing $\lambda$ and pairing gap.
Although the pair correlation could diminish
for small $\lambda$~\cite{ref:HM04},
the unpaired-particle haloing starts bringing into action earlier,
at sizable $|\lambda|$.
Whereas similar broadening mechanism for excited states was pointed out
in Ref.~\cite{ref:Yam05},
it may be responsible for density distributions of g.s. in odd-$N$ nuclei.

Role of the pair correlation in the unpaired-particle haloing
will be further clarified
if we use the HF+BCS scheme as an approximation of the HFB~\cite{can-bas}.
In the HF+BCS, the q.p. energy is expressed
by $\varepsilon_k=\sqrt{(\epsilon^\mathrm{HF}_k-\lambda)^2+\Delta_k^2}$,
where $\epsilon^\mathrm{HF}_k$ is the s.p. energy in the HF
and $\Delta_k$ is the pairing gap.
We reasonably assume $\varepsilon^\mathrm{min}=\varepsilon_{k_1}
\approx|\Delta_{k_1}|$
because $\epsilon^\mathrm{HF}_{k_1}\approx\lambda$,
yielding $\eta^\mathrm{min}_\pm\approx\sqrt{2M(|\epsilon^\mathrm{HF}_{k_1}|
  \pm|\Delta_{k_1}|)}$.
Compared with the asymptotics in the HF,
which corresponds to the $\Delta_{k_1}\to 0$ limit,
$\rho(\mathbf{r})$ damps more slowly for increasing $r$
with Eq.~(\ref{eq:dns-asymp2})
while more quickly with Eq.~(\ref{eq:dns-asymp}).
Thus, opposite to the pairing anti-halo effect in the even-$N$ cases,
the pair correlation may enhance a halo when an unpaired particle is present.
The asymptotics of Eq.~(\ref{eq:dns-asymp2}) is in harmony
with the energy of the q.p. state $-(|\lambda|-\varepsilon^\mathrm{min})
=\lambda+\varepsilon_{k_1}\approx\epsilon^\mathrm{HF}_{k_1}+|\Delta_{k_1}|$.
Therefore, it can be interpreted
as the pairing leads to loose binding of q.p. states
and it gives rise to the unpaired-particle haloing.


\noindent\textit{Results.}

It has been disclosed via $\sigma_R$'s~\cite{ref:Mg37exp}
that the rms matter radii $\sqrt{\langle r^2\rangle}$
in the neutron-rich isotopes $^{34-38}$Mg
have quite irregular dependence on $N$;
$\sqrt{\langle r^2\rangle}(\mbox{$^{35}$Mg})$ seems smaller
than the average of the neighboring even-$N$ nuclei,
\textit{i.e.} $\big[\sqrt{\langle r^2\rangle}(\mbox{$^{34}$Mg})
  +\sqrt{\langle r^2\rangle}(\mbox{$^{36}$Mg})\big]/2$,
whereas $\sqrt{\langle r^2\rangle}(\mbox{$^{37}$Mg})$
is clearly enhanced~\cite{ref:Wat14}.
This enhancement of $\sqrt{\langle r^2\rangle}(\mbox{$^{37}$Mg})$
indicates a halo formed by a $p$-wave neutron,
as experimentally confirmed in Ref.~\cite{ref:Mg37exp2}.
In Refs.~\cite{ref:Mg37exp,ref:Wat14},
matter radii of a long chain of the Mg isotopes
calculated within the antisymmetrized molecular dynamics (AMD) were shown,
in which the Gogny-D1S interaction~\cite{ref:D1S} was adopted.
Although the AMD calculation successfully reproduces overall trend,
it fails to describe the $N$-dependence in $^{34-38}$Mg.
The discrepancy in $^{37}$Mg seems to support its halo nature,
because the AMD wave functions in Refs.~\cite{ref:Mg37exp,ref:Wat14} contain
no long-tailed components.
In Ref.~\cite{ref:UHS17}, the staggering of $\sigma_R$'s in $^{36-38}$Mg
was investigated via the HFB calculation,
but on top of the deformed WS potential.
The staggering was accounted for as the pairing anti-halo effect in $^{38}$Mg,
while the halo in $^{37}$Mg is attributed to the conventional s.p. picture
with quenched pair correlation.

\begin{figure}
  \includegraphics[scale=0.5]{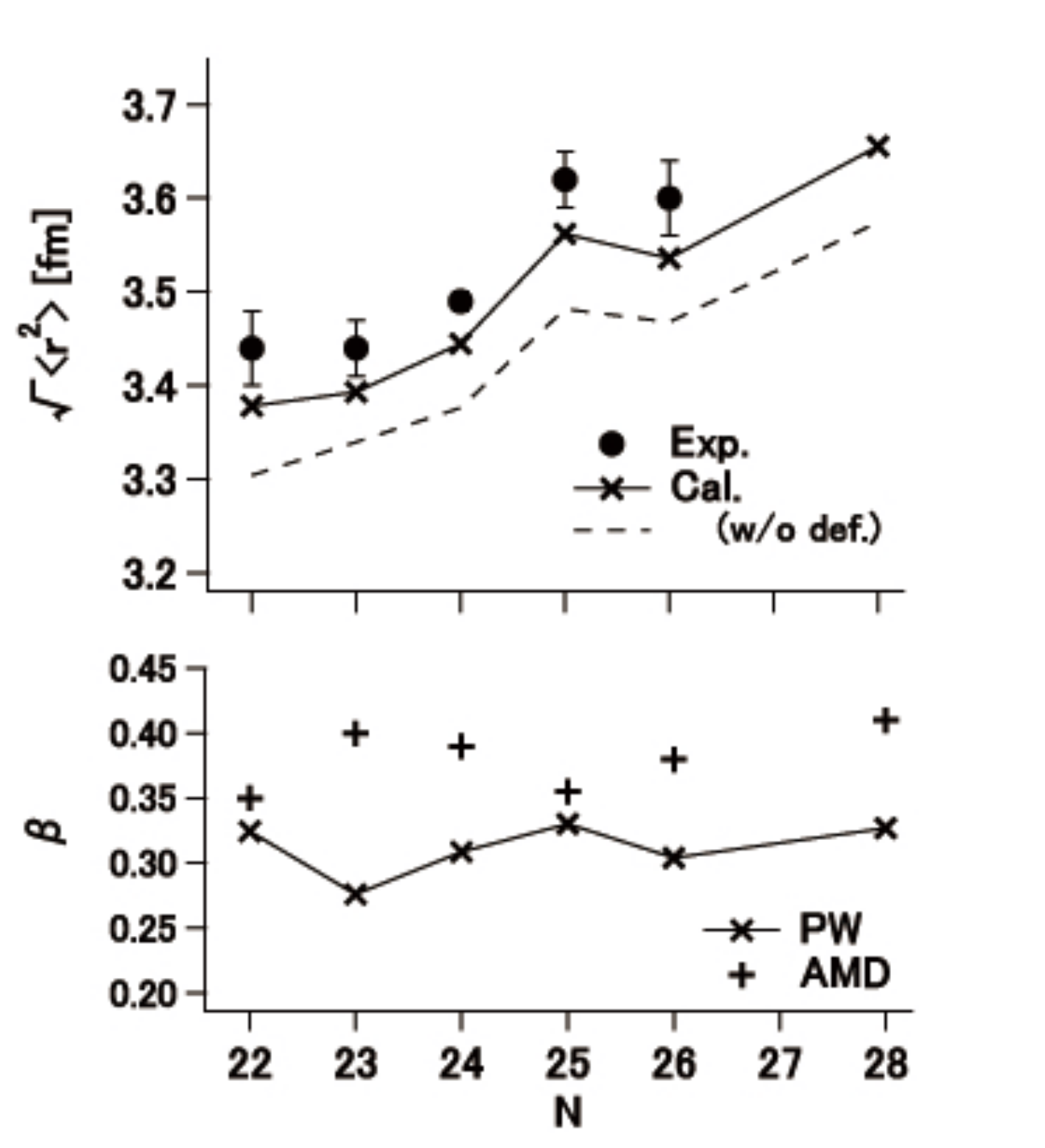}\vspace*{3mm}
  \caption{
    Upper panel: Rms matter radii $\sqrt{\langle r^2\rangle}$ in $^{34-40}$Mg.
    The crosses connected by the solid line
    represent the HFB results with M3Y-P6,
    and the dots with error bars are experimental values
    extracted from $\sigma_R$'s~\cite{ref:Wat14}.
    For reference, $\bar{r}_0$ values [see Eq.~(\ref{eq:rbar-beta})]
    are plotted by the dashed line.\\
    Lower panel: Deformation parameter $\beta$.
    The crosses are obtained from the present HFB results with M3Y-P6
    via Eq.~(\ref{eq:rbar-beta}).
    The pluses are the AMD results quoted from Ref.~\cite{ref:Wat14}.
\label{fig:Mg_rad}}
\end{figure}

Self-consistent HFB calculations with the M3Y-P6 interaction
have been implemented in the present work.
The calculated rms matter radii in $^{34-40}$Mg are shown
in Fig.~\ref{fig:Mg_rad},
in comparison with the experimental values
extracted from $\sigma_R$'s~\cite{ref:Wat14}.
No data are available for $^{40}$Mg,
and $^{39}$Mg is predicted to be unbound.
It is found that the present calculations
excellently reproduce the $N$-dependence of $\sqrt{\langle r^2\rangle}$
in $^{34-38}$Mg,
though the absolute values are slightly underestimated.

\begin{figure}
  \hspace*{-1cm}
  \includegraphics[scale=0.4]{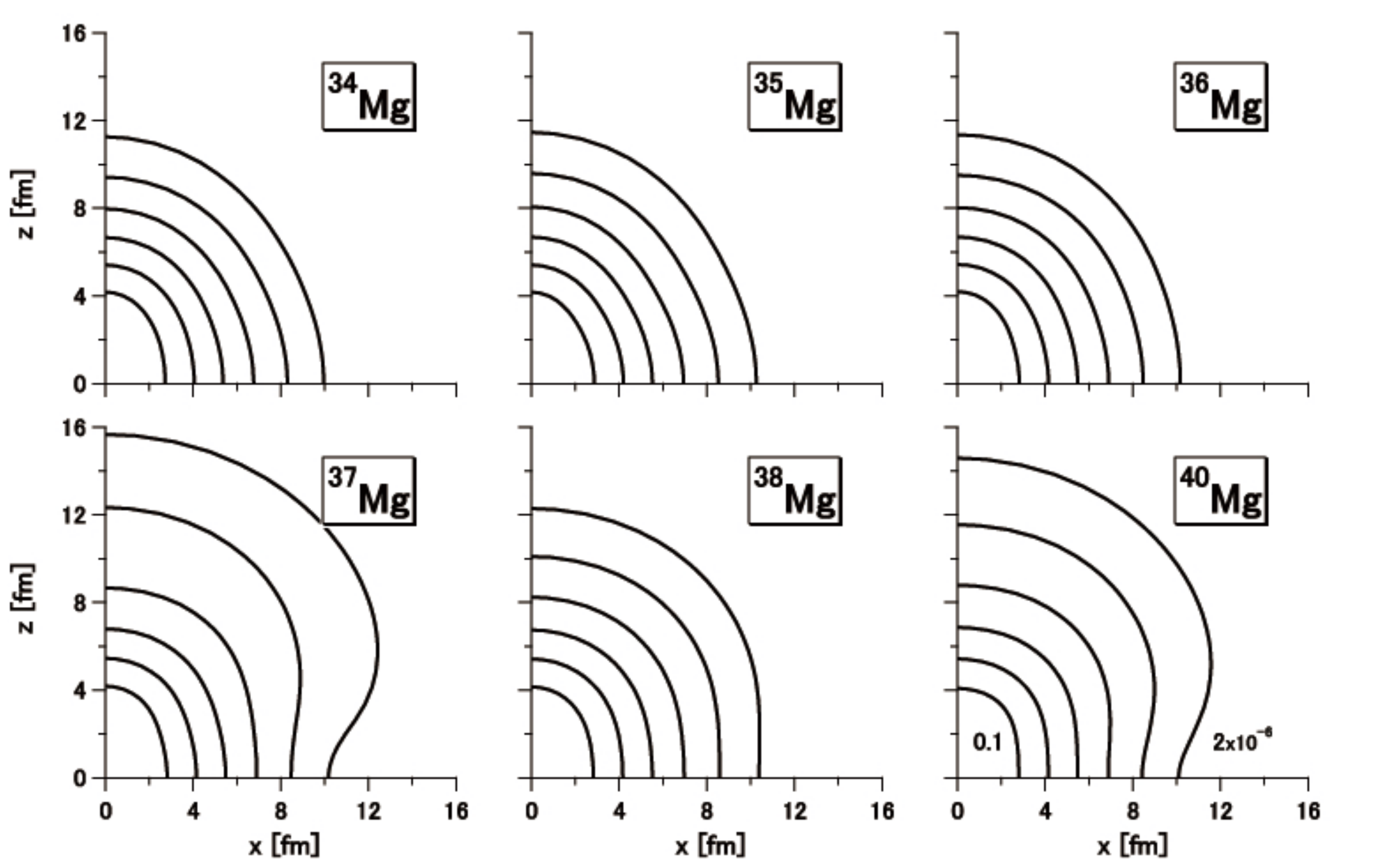}\vspace*{3mm}
  \caption{Contour plot of $\rho(\mathbf{r})$ on the $zx$-plane for $^{34-40}$Mg,
    obtained by the HFB calculations with the M3Y-P6 interaction.
    Positions of $\rho(\mathbf{r})=0.1$, $2\times 10^{-2}$, $2\times 10^{-3}$,
    $2\times 10^{-4}$, $2\times 10^{-5}$ and $2\times 10^{-6}\,\mathrm{fm}^{-3}$
    are presented.
\label{fig:Mg_rhocnt}}
\end{figure}

To examine relevance of neutron halos,
we depict the calculated density distributions
in terms of the equi-density lines in the $zx$-plane.
The $z$-axis is taken to be the symmetry axis here,
and the $x$ coordinate in the figure represents the distance from the $z$-axis.
In addition to the symmetry about the rotation around the $z$-axis,
we have the reflection symmetry with respect to the $xy$-plane.
The equi-density lines are drawn for exponentially decreasing values
of $\rho(\mathbf{r})$,
except the highest value $0.1\,\mathrm{fm}^{-3}$.
The almost constant interval of the lines for large $r\,(=\sqrt{x^2+z^2})$
implies that the exponential asymptotics
as in Eqs.~(\ref{eq:dns-asymp},\ref{eq:dns-asymp2})
are well described.
It is obvious from Fig.~\ref{fig:Mg_rhocnt}
that the present calculation predicts halos in $^{37}$Mg
and $^{40}$Mg.
These halos have peanut-shape in the intrinsic states,
as a result of the $p$-wave contribution.
In practice, the unpaired particle in $^{37}$Mg occupies
an $\Omega^\pi=(1/2)^-$ level,
$[N\,n_3\,\Lambda\,\Omega]=[3\,1\,0\,\frac{1}{2}]$
in terms of the Nilsson asymptotic quantum number,
which consists mainly of the $p_{3/2}$ component.
The shape of the halo in $^{40}$Mg is analogous,
as will be discussed later.

In the result of Ref.~\cite{ref:UHS17},
the g.s. of $^{37}$Mg stayed in the normal fluid phase.
In contrast, $^{37}$Mg contains pair correlation in the present result,
having the neutron pair energy of $\sim 3\,\mathrm{MeV}$.
Moreover, the neutron chemical potential is $-2.25\mathrm{MeV}$,
not small enough to account for the halo structure
if the pair correlation is fully ignored.
However, the pairing in $^{37}$Mg enhances a halo
through the unpaired-particle haloing, not preventing it.
We now have $\varepsilon^\mathrm{min}=2.02\,\mathrm{MeV}$.
If $\rho(\mathbf{r})$ followed the asymptotics
with $\eta^\mathrm{min}_+$ [Eq.~(\ref{eq:dns-asymp})],
no halo should be produced
because $|\lambda|+\varepsilon^\mathrm{min}\approx 4.3\,\mathrm{MeV}$.
On the other hand, the unpaired-particle haloing,
\textit{i.e.} the asymptotics with $\eta^\mathrm{min}_-$ [Eq.~(\ref{eq:dns-asymp2})],
well accounts for the broad density distribution,
with $|\lambda|-\varepsilon^\mathrm{min}\approx 0.2\,\mathrm{MeV}$.

In the present calculations,
deformation is determined self-consistently, depending on $N$.
It deserves noting that the deformation parameter $\beta$,
which has been extracted through Eq.~(\ref{eq:rbar-beta}),
is larger in $^{37}$Mg than in the neighboring isotopes.
This larger $\beta$ might not look essential
in the enhancement of $\sqrt{\langle r^2\rangle}(\mbox{$^{37}$Mg})$
since the larger $\bar{r}_0(\mbox{$^{37}$Mg})$ already accounts for
most of the enhancement in Fig.~\ref{fig:Mg_rad}.
However, it is noticed that $\bar{r}_0$ and $\beta$ contribute cooperatively
to the enhancement.
In order for the last neutron
to occupy the $\Omega^\pi=(1/2)^-$ level in $^{37}$Mg,
there should be a level crossing between this level and
the $\Omega^\pi=(5/2)^-$ level dominated by the $0f_{7/2}$ component
(\textit{i.e.} $[3\,1\,2\,\frac{5}{2}]$),
on the prolate side.
As discussed in Ref.~\cite{ref:UHS17},
relatively large deformation is needed for this crossing.
In $^{36,38}$Mg,
the occupation probability on the $\Omega^\pi=(1/2)^-$ q.p. level is lower
than that on the $\Omega^\pi=(5/2)^-$ level.
This indicates that the larger deformation in $^{37}$Mg
is important for the one q.p. state with $\Omega^\pi=(1/2)^-$ to be its g.s.,
by which the large $\bar{r}_0$ and the halo become possible.
In this respect, the deformation assists the unpaired-particle haloing
to operate.
Conversely, the halo drives the larger deformation so as to gain energy.

As well as the enhancement of $\sqrt{\langle r^2\rangle}(\mbox{$^{37}$Mg})$,
the present calculation reproduces
reduction of $\sqrt{\langle r^2\rangle}(\mbox{$^{35}$Mg})$.
Because there is no visible reduction of $\bar{r}_0$,
this is attributed to the smaller $\beta$ in $^{35}$Mg than in $^{34,36}$Mg,
as shown in the lower panel of Fig.~\ref{fig:Mg_rad}.
Note that such $N$-dependence of the deformation is hard to be realized
without self-consistent calculations.
It is commented that the $\beta$ values in the AMD results
in Ref.~\cite{ref:Wat14} have quite different $N$-dependence,
which seems a source of the discrepancy
in $\sqrt{\langle r^2\rangle}(\mbox{$^{34-38}$Mg})$.

For $^{40}$Mg, the pair correlation is quenched.
Therefore, this nucleus is free from the pairing anti-halo effect
and from influence of the continuum.
The highest occupied s.p. level has $\Omega^\pi=(1/2)^-$
dominated by the $p$-wave components,
corresponding to $[3\,1\,0\,\frac{1}{2}]$ again
and accounting for the peanut-shape halo in Fig.~\ref{fig:Mg_rhocnt}.
This result is consistent with the HFB result with the Gogny-D1S interaction
in Ref.~\cite{ref:Nak08}.

It should be mentioned that,
based on the relativistic Hartree-Bogolyubov calculations,
neutron halos up to more neutron-rich Mg isotopes ($^{42-46}$Mg)
have been argued,
though restricted to even-$N$~\cite{ref:ZMRZ10,ref:CRM14,ref:Meng06,ref:Li12}.
We here note that the Mg nuclei beyond $N=28$ are not bound
in the present calculation using the M3Y-P6 interaction,
as in the HFB calculations with the Gogny-D1S interactions~\cite{ref:D1S-Web}.
In Fig.~1(b) of Ref.~\cite{ref:ZMRZ10},
we find peanut-shape for the predicted halo in $^{44}$Mg,
as a result of the $p$-wave dominance.
Another notable point for $^{44}$Mg in Ref.~\cite{ref:ZMRZ10} would be that,
while the core is deformed with prolate shape,
the halo has oblate shape.
This is traced back to the $\Lambda$ value (\textit{i.e.} the $z$-component
of the orbital angular momentum) of the halo orbitals,
in addition to the $p$-wave dominance;
deformation of the halo depends on quantum numbers
of the s.p. orbits~\cite{ref:MNA97,ref:ZMRZ10}.
In the present calculations for $^{37,40}$Mg,
the halo orbit is predominantly comprised of the $\Lambda=0$ component
(see discussion in Ref.~\cite{ref:Nak08} for $^{40}$Mg),
yielding prolate deformation as the core.

\noindent\textit{Summary and outlook.}

The irregular $N$-dependence of the matter radii in $^{34-38}$Mg
has been investigated via self-consistent axial HFB calculations
with the semi-realistic M3Y-P6 interaction.
The staggering is reproduced excellently,
in which the pairing and the deformation affect in an intertwined manner.
The results do not indicate halos in $^{34-36,38}$Mg,
while deformed halos with the peanut-shape are predicted for $^{37,40}$Mg.
The deformation is reduced in $^{35}$Mg,
which yields the smaller $\sqrt{\langle r^2\rangle}(\mbox{$^{35}$Mg})$
compared to the average of $\sqrt{\langle r^2\rangle}(\mbox{$^{34}$Mg})$
and $\sqrt{\langle r^2\rangle}(\mbox{$^{36}$Mg})$.
In relevance to the halo in $^{37}$Mg, we point out a new mechanism,
called unpaired-particle haloing,
that the pairing can enhance halos in odd-$N$ nuclei.
This is because of the asymptotics of the last unpaired neutron.
It should also be stressed that the $N$-dependence of the deformation
assists the halo in $^{37}$Mg.

Although extensive investigation of lighter Mg nuclei
and of the staggering of the matter radii
in $^{30-32}$Ne~\cite{ref:Ne29-31exp,ref:Ne31exp2} will be interesting,
we leave it as a future subject.
Connected to the loss of magicity,
careful study is needed for the Ne and Mg nuclei near $N=20$,
in which spherical-deformed shape coexistence may occur~\cite{ref:SNM16}
and rotational correlations could be
of particular significance~\cite{ref:RER02}.
Another topic will be influence of the predicted exotic peanut-shape
of the halos on observables.
This is far from trivial and waits for further discussions.

Nuclear deformation, pairing and halos take place at the g.s.
so as to lower the energy.
For nuclei sufficiently close to the drip line,
loosely bound nucleons sometimes form halos
so that the kinetic energy (including contribution of the centrifugal potential)
could be small.
Deformation and pairing act cooperatively if possible,
as exemplified by $^{37}$Mg
and it is naturally expected that effects of deformation and pairing
are intertwined in halo nuclei.
Assisted by these effects, the unpaired-particle haloing
implies that odd-$N$ nuclei tend to have larger radii
than their neighboring even-$N$ nuclei near the drip line, if they are bound.
This argument for the even-odd effect of the nuclear radii will apply
to a wide range of the nuclear masses,
and seems compatible with many experimental data.

\begin{acknowledgments}
%
One of the authors (H.N.) is grateful to M.K. Gaidarov,
J. Meng and S.-G. Zhou for comments on the manuscript.
This work is financially supported in part
by the JSPS KAKENHI with Grant Number~16K05342.
A part of the numerical calculations have been performed on HITAC SR24000
at Institute of Management and Information Technologies in Chiba University.
\end{acknowledgments}


\begin{thebibliography}{99}
\bibitem{ref:Rii94} K. Riisager, Rev. Mod. Phys. \textbf{66}, 1105 (1994)
\bibitem{ref:Tan85} I. Tanihata, H. Hamagaki, O. Hashimoto, Y. Shida,
  N. Yoshikawa, K. Sugimoto, O. Yamakawa, T. Kobayashi and N. Takahashi,
  Phys. Rev. Lett. \textbf{55}, 2676 (1985).
\bibitem{ref:Mg37exp} M. Takechi, \textit{et al.},
  Phys. Rev. C \textbf{90}, 061305(R) (2014).
\bibitem{ref:Mg37exp2} N. Kobayashi, \textit{et al.},
  Phys. Rev. Lett. \textbf{112}, 242501 (2014).
\bibitem{ref:Kar75} P.J. Karol, Phys. Rev. C \textbf{11}, 1203 (1975).
\bibitem{ref:Wat14} S. Watanabe, \textit{et al.},
  Phys. Rev. C \textbf{89}, 044610 (2014).
\bibitem{ref:MNA97} T. Misu, W. Nazarewicz and S. \AA berg,
  Nucl. Phys. A \textbf{614}, 44 (1997).
\bibitem{ref:Nak08} H. Nakada, Nucl. Phys. A \textbf{808}, 47 (2008).
\bibitem{ref:UHS17} Y. Urata, K. Hagino and H. Sagawa,
  Phys. Rev. C \textbf{96}, 064311 (2017).
\bibitem{ref:BDP00} K. Bennaceur, J. Dobaczewski and M. P{\l}oszajczak,
  Phys. Lett. B \textbf{496} (2000) 154.
\bibitem{ref:ZMRZ10} S.-G. Zhou, J. Meng, P. Ring and E.-G. Zhao,
  Phys. Rev. C \textbf{82}, 011301(R) (2010).
\bibitem{ref:GSAM14} M.K. Gaidarov, P. Sarriguren, A.N. Antonov
  and E. Moya de Guerra, Phys. Rev. C \textbf{89}, 064301 (2014).
\bibitem{ref:CRM14} Y. Chen, P. Ring and J. Meng,
  Phys. Rev. C \textbf{89}, 014312 (2014).
\bibitem{ref:ZCMR17} Y. Zhang, Y. Chen, J. Meng and P. Ring,
  Phys. Rev. C \textbf{95}, 014316 (2017).
\bibitem{ref:MR96} J. Meng and P. Ring,
  Phys. Rev. Lett. \textbf{77}, 3963 (1996).
\bibitem{ref:Nak13} H. Nakada, Phys. Rev. C \textbf{87}, 014336 (2013).
\bibitem{ref:Nak03} H. Nakada, Phys. Rev. C \textbf{68}, 014316 (2003).
\bibitem{ref:SNM16} Y. Suzuki, H. Nakada and S. Miyahara,
  Phys. Rev. C \textbf{94}, 024343 (2016).
\bibitem{ref:NS02} H. Nakada and M. Sato,
  Nucl. Phys. A \textbf{699}, 511 (2002);
  \textit{ibid.} \textbf{714}, 696 (2003).
\bibitem{ref:Nak06} H. Nakada, Nucl. Phys. A \textbf{764}, 117 (2006);
  \textit{ibid.} \textbf{801}, 169 (2008).
\bibitem{ref:NMYM09} H. Nakada, K. Mizuyama, M. Yamagami and M. Matsuo,
  Nucl. Phys. A \textbf{828}, 283 (2009).
\bibitem{ref:NSM13} H. Nakada, K. Sugiura and J. Margueron,
 Phys. Rev. C \textbf{87}, 067305 (2013).
\bibitem{ref:NS14} H. Nakada and K. Sugiura,
  Prog. Theor. Exp. Phys. \textbf{2014}, 033D02.
\bibitem{ref:RS80} P. Ring and P. Schuck,
  \textit{The Nuclear Many-Body Problem} (Springer-Verlag, New York, 1980).
\bibitem{ref:EFA} S. Perez-Martin and L.M. Robledo,
  Phys. Rev. C \textbf{78}, 014304 (2008);
  N. Schunck, J. Dobaczewski, J. McDonnell, J. More, W. Nazarewicz,
  J. Sarich and M.V. Stoitsov, Phys. Rev. C \textbf{81}, 024316 (2010).
\bibitem{ref:Ham07} I. Hamamoto, Phys. Rev. C \textbf{76}, 054319 (2007).
\bibitem{ref:BM1} A. Bohr and B.R. Mottelson,
  \textit{Nuclear Structure}, vol.~1 (Benjamin, New York, 1969).
\bibitem{ref:SkP} J. Dobaczewski, H. Flocard and J. Treiner,
  Nucl. Phys. A \textbf{422}, 103 (1984).
\bibitem{ref:RJM92} K. Riisager, A.S. Jensen and P. M\o ller,
  Nucl. Phys. A \textbf{548}, 393 (1992).
\bibitem{ref:HM04} I. Hamamoto, Phys. Rev. C \textbf{69}, 041306(R) (2004);
  I. Hamamoto and B.R. Mottelson,
  Phys. Rev. C \textbf{69}, 064302 (2004);
  I. Hamamoto and H. Sagawa, Phys. Rev. C \textbf{70}, 034317 (2004).
\bibitem{ref:Yam05} M. Yamagami, Phys. Rev. C \textbf{72}, 064308 (2005).
\bibitem{can-bas} Although the HF+BCS scheme does not give correct asymptotics
  as typically known as the neutron-gas problem,
  it is here used only for assessing the q.p. energy $\varepsilon_k$.
  As in Ref.~\cite{ref:BDP00}, a similar argument is applicable
  with the canonical-basis representation of the HFB
  under a certain approximation.
\bibitem{ref:D1S} J.F. Berger, M. Girod and D. Gogny,
 Comp. Phys. Comm. \textbf{63}, 365 (1991).
\bibitem{ref:Meng06} J. Meng, H. Toki, S.G. Zhou, S.Q. Zhang, W.H. Long
  and L.S. Geng, Prog. Part. Nucl. Phys. \textbf{57}, 470 (2006);
  J. Meng and S.G. Zhou, J. Phys. G \textbf{42}, 093101 (2015).
\bibitem{ref:Li12} L. Li, J. Meng, P. Ring, E.-G. Zhao and S.-G. Zhou,
  Phys. Rev. C \textbf{85}, 024312 (2012);
\bibitem{ref:D1S-Web} S. Hilaire and M. Girod,
 http://www-phynu.cea.fr/science\_en\_ligne/carte\_potentiels\_microscopiques/
 carte\_potentiel\_nucleaire\_eng.htm.
\bibitem{ref:Ne29-31exp} M. Takechi, \textit{et al.},
  Phys. Lett. B \textbf{707}, 357 (2012).
\bibitem{ref:Ne31exp2} T. Nakamura, \textit{et al.},
  Phys. Rev. Lett. \textbf{112}, 142501 (2014).
\bibitem{ref:RER02} R. Rodor\'{i}guez-Guzm\'{a}n, J.L. Egido and L.M. Robledo,
  Nucl. Phys. A \textbf{709}, 201 (2002);
  M. Borrajo and J.L. Egido, Phys. Lett. B \textbf{764}, 328 (2017).
\end{thebibliography}

\end{document}